\input harvmac
\def\ev#1{\langle#1\rangle}
\input amssym
\input epsf

\def\unit{\relax{\rm 1\kern-.26em I}}
\def\nada{\relax{\rm 0\kern-.30em l}}
\def\tilde{\widetilde}



\def\det{{\rm det}}

\noblackbox
\def\IL{\relax{\rm I\kern-.18em L}}
\def\IH{\relax{\rm I\kern-.18em H}}
\def\IR{\relax{\rm I\kern-.18em R}}
\def\IC{\relax\hbox{$\inbar\kern-.3em{\rm C}$}}
\def\IZ{\relax\ifmmode\mathchoice
{\hbox{\cmss Z\kern-.4em Z}}{\hbox{\cmss Z\kern-.4em Z}}
{\lower.9pt\hbox{\cmsss Z\kern-.4em Z}} {\lower1.2pt\hbox{\cmsss
Z\kern-.4em Z}}\else{\cmss Z\kern-.4em Z}\fi}
\def\CM {{\cal M}}

\def\CO {{\cal O}}

\def\CM {{\cal M}}

\def\CO {{\cal O}}

\def\det{{\rm det}}
\def\Tr{{\rm Tr}}

\font\manual=manfnt \def\dbend{\lower3.5pt\hbox{\manual\char127}}

\def\IZ{\relax\ifmmode\mathchoice
{\hbox{\cmss Z\kern-.4em Z}}{\hbox{\cmss Z\kern-.4em Z}}
{\lower.9pt\hbox{\cmsss Z\kern-.4em Z}} {\lower1.2pt\hbox{\cmsss
Z\kern-.4em Z}}\else{\cmss Z\kern-.4em Z}\fi}
\def\half {{1\over 2}}

\def\lfm#1{\medskip\noindent\item{#1}}

\def\bar{\overline}

\def\rt2{\sqrt{2}}
\def\irt2{{1\over\sqrt{2}}}

\def\slashchar#1{\setbox0=\hbox{$#1$}           
   \dimen0=\wd0                                 
   \setbox1=\hbox{/} \dimen1=\wd1               
   \ifdim\dimen0>\dimen1                        
      \rlap{\hbox to \dimen0{\hfil/\hfil}}      
      #1                                        
   \else                                        
      \rlap{\hbox to \dimen1{\hfil$#1$\hfil}}   
      /                                         
   \fi}

\def\foursqr#1#2{{\vcenter{\vbox{
    \hrule height.#2pt
    \hbox{\vrule width.#2pt height#1pt \kern#1pt
    \vrule width.#2pt}
    \hrule height.#2pt
    \hrule height.#2pt
    \hbox{\vrule width.#2pt height#1pt \kern#1pt
    \vrule width.#2pt}
    \hrule height.#2pt
        \hrule height.#2pt
    \hbox{\vrule width.#2pt height#1pt \kern#1pt
    \vrule width.#2pt}
    \hrule height.#2pt
        \hrule height.#2pt
    \hbox{\vrule width.#2pt height#1pt \kern#1pt
    \vrule width.#2pt}
    \hrule height.#2pt}}}}
\def\psqr#1#2{{\vcenter{\vbox{\hrule height.#2pt
    \hbox{\vrule width.#2pt height#1pt \kern#1pt
    \vrule width.#2pt}
    \hrule height.#2pt \hrule height.#2pt
    \hbox{\vrule width.#2pt height#1pt \kern#1pt
    \vrule width.#2pt}
    \hrule height.#2pt}}}}
\def\sqr#1#2{{\vcenter{\vbox{\hrule height.#2pt
    \hbox{\vrule width.#2pt height#1pt \kern#1pt
    \vrule width.#2pt}
    \hrule height.#2pt}}}}

\lref\HananyNA{
  A.~Hanany and Y.~Oz,
  ``On the quantum moduli space of vacua of N=2 supersymmetric SU(N(c)) gauge
  theories,''
  Nucl.\ Phys.\ B {\bf 452}, 283 (1995)
  [arXiv:hep-th/9505075].
}

\lref\IntriligatorCP{
  K.~Intriligator and N.~Seiberg,
  ``Lectures on supersymmetry breaking,''
  arXiv:hep-ph/0702069.
}

\lref\DineYW{
  M.~Dine and A.~E.~Nelson,
  ``Dynamical supersymmetry breaking at low-energies,''
  Phys.\ Rev.\ D {\bf 48}, 1277 (1993)
  [arXiv:hep-ph/9303230].
}

\lref\DimopoulosWW{
  S.~Dimopoulos, G.~R.~Dvali, R.~Rattazzi and G.~F.~Giudice,
  ``Dynamical soft terms with unbroken supersymmetry,''
  Nucl.\ Phys.\  B {\bf 510}, 12 (1998)
  [arXiv:hep-ph/9705307].
}

\lref\ArgyresCB{
  P.~C.~Argyres and M.~Edalati,
  ``Generalized Konishi anomaly, Seiberg duality and singular effective
  superpotentials,''
  arXiv:hep-th/0511272.
}
\lref\DineGM{
  M.~Dine, J.~L.~Feng and E.~Silverstein,
  ``Retrofitting O'Raifeartaigh models with dynamical scales,''
  Phys.\ Rev.\ D {\bf 74}, 095012 (2006)
  [arXiv:hep-th/0608159].
}
\lref\AharonyMY{
 O.~Aharony and N.~Seiberg,
 ``Naturalized and simplified gauge mediation,''
 arXiv:hep-ph/0612308.
}

\lref\WittenIH{
  E.~Witten,
  ``Mass Hierarchies In Supersymmetric Theories,''
  Phys.\ Lett.\ B {\bf 105}, 267 (1981).
}
\lref\DimopoulosGM{
  S.~Dimopoulos and S.~Raby,
  ``Geometric Hierarchy,''
  Nucl.\ Phys.\  B {\bf 219}, 479 (1983).
}

\lref\NappiHM{
  C.~R.~Nappi and B.~A.~Ovrut,
  ``Supersymmetric Extension Of The SU(3) X SU(2) X U(1) Model,''
  Phys.\ Lett.\  B {\bf 113}, 175 (1982).
}

\lref\CsakiWI{
  C.~Csaki, Y.~Shirman and J.~Terning,
  ``A simple model of low-scale direct gauge mediation,''
  arXiv:hep-ph/0612241.
}

\lref\BanksMG{
  T.~Banks and V.~Kaplunovsky,
  ``Nosonomy Of An Upside Down Hierarchy Model. 1,''
  Nucl.\ Phys.\  B {\bf 211}, 529 (1983).
}
\lref\KaplunovskyYX{
  V.~Kaplunovsky,
  ``Nosonomy Of An Upside Down Hierarchy Model. 2,''
  Nucl.\ Phys.\  B {\bf 233}, 336 (1984).
}
\lref\tyui{A.~Hanany and Y.~Oz, ``On the quantum moduli space of
vacua of N=2 supersymmetric SU(N(c)) gauge theories,'' Nucl.\
Phys.\ B {\bf 452}, 283 (1995) [arXiv:hep-th/9505075].
}
\lref\tyuii{P.~C.~Argyres, M.~R.~Plesser and A.~D.~Shapere, ``The
Coulomb phase of N=2 supersymmetric QCD,'' Phys.\ Rev.\ Lett.\
{\bf 75}, 1699 (1995) [arXiv:hep-th/9505100].
}
\lref\tyuiii{J.~A.~Minahan and D.~Nemeschansky, ``Hyperelliptic
curves for supersymmetric Yang-Mills,'' Nucl.\ Phys.\ B {\bf 464},
3 (1996) [arXiv:hep-th/9507032].
}
\lref\tyuiv{I.~M.~Krichever and D.~H.~Phong, ``On the integrable
geometry of soliton equations and N = 2  supersymmetric gauge
theories,'' J.\ Diff.\ Geom.\  {\bf 45}, 349 (1997)
[arXiv:hep-th/9604199].
}
\lref\tyuv{E.~D'Hoker, I.~M.~Krichever and D.~H.~Phong, ``The
effective prepotential of N = 2 supersymmetric SU(N(c)) gauge
theories,'' Nucl.\ Phys.\ B {\bf 489}, 179 (1997)
[arXiv:hep-th/9609041].
}
\lref\PreskillMZ{
  J.~Preskill,
  ``Subgroup Alignment In Hypercolor Theories,''
  Nucl.\ Phys.\ B {\bf 177}, 21 (1981).
}
\lref\ISS{
  K.~A.~Intriligator, N.~Seiberg and S.~H.~Shenker,
  ``Proposal for a simple model of dynamical SUSY breaking,''
  Phys.\ Lett.\ B {\bf 342}, 152 (1995)
  [arXiv:hep-ph/9410203].
}
\lref\DineYW{
  M.~Dine and A.~E.~Nelson,
  ``Dynamical supersymmetry breaking at low-energies,''
  Phys.\ Rev.\ D {\bf 48}, 1277 (1993)
  [arXiv:hep-ph/9303230].
}
\lref\AffleckUZ{
  I.~Affleck, M.~Dine and N.~Seiberg,
  ``Calculable Nonperturbative Supersymmetry Breaking,''
  Phys.\ Rev.\ Lett.\  {\bf 52}, 1677 (1984).
}
\lref\IntriligatorCP{
  K.~Intriligator and N.~Seiberg,
  ``Lectures on supersymmetry breaking,''
  arXiv:hep-ph/0702069.
}
\lref\AffleckXZ{
  I.~Affleck, M.~Dine and N.~Seiberg,
  ``Dynamical Supersymmetry Breaking In Four-Dimensions And Its
  Phenomenological Implications,''
  Nucl.\ Phys.\ B {\bf 256}, 557 (1985).
}
\lref\GiudiceBP{
  G.~F.~Giudice and R.~Rattazzi,
  ``Theories with gauge-mediated supersymmetry breaking,''
  Phys.\ Rept.\  {\bf 322}, 419 (1999)
  [arXiv:hep-ph/9801271].
}
\lref\WittenKV{
  E.~Witten,
  ``Mass Hierarchies In Supersymmetric Theories,''
  Phys.\ Lett.\ B {\bf 105}, 267 (1981).
}
\lref\LutyVR{
  M.~A.~Luty and J.~Terning,
  ``Improved single sector supersymmetry breaking,''
  Phys.\ Rev.\ D {\bf 62}, 075006 (2000)
  [arXiv:hep-ph/9812290].
}

\lref\BaggerHH{
  J.~Bagger, E.~Poppitz and L.~Randall,
  ``The R axion from dynamical supersymmetry breaking,''
  Nucl.\ Phys.\ B {\bf 426}, 3 (1994)
  [arXiv:hep-ph/9405345].
}
\lref\WessCP{
  J.~Wess and J.~Bagger,
  ``Supersymmetry and supergravity,''
}
\lref\NSd{
  N.~Seiberg,
  ``Electric - magnetic duality in supersymmetric nonAbelian gauge theories,''
  Nucl.\ Phys.\ B {\bf 435}, 129 (1995)
  [arXiv:hep-th/9411149].
}
\lref\ArkaniHamedFQ{
  N.~Arkani-Hamed, M.~A.~Luty and J.~Terning,
  ``Composite quarks and leptons from dynamical supersymmetry breaking  without
  messengers,''
  Phys.\ Rev.\ D {\bf 58}, 015004 (1998)
  [arXiv:hep-ph/9712389].
}
\lref\WittenDF{
  E.~Witten,
  ``Constraints On Supersymmetry Breaking,''
  Nucl.\ Phys.\ B {\bf 202}, 253 (1982).
}

\lref\AffleckMK{
  I.~Affleck, M.~Dine and N.~Seiberg,
  ``Dynamical Supersymmetry Breaking In Supersymmetric QCD,''
  Nucl.\ Phys.\ B {\bf 241}, 493 (1984).
}

\lref\AffleckXZ{
  I.~Affleck, M.~Dine and N.~Seiberg,
  ``Dynamical Supersymmetry Breaking In Four-Dimensions And Its
  Phenomenological Implications,''
  Nucl.\ Phys.\ B {\bf 256}, 557 (1985).
}
\lref\CachazoJY{
  F.~Cachazo, K.~A.~Intriligator and C.~Vafa,
  ``A large N duality via a geometric transition,''
  Nucl.\ Phys.\ B {\bf 603}, 3 (2001)
  [arXiv:hep-th/0103067].
}
\lref\CachazoRY{
  F.~Cachazo, M.~R.~Douglas, N.~Seiberg and E.~Witten,
  ``Chiral rings and anomalies in supersymmetric gauge theory,''
  JHEP {\bf 0212}, 071 (2002)
  [arXiv:hep-th/0211170].
}
\lref\BanksRU{
  T.~Banks and M.~Johnson,
  ``Regulating eternal inflation,''
  arXiv:hep-th/0512141.
}

\lref\DijkgraafDH{
  R.~Dijkgraaf and C.~Vafa,
  ``A perturbative window into non-perturbative physics,''
  arXiv:hep-th/0208048.
}
\lref\BaggerHH{
  J.~Bagger, E.~Poppitz and L.~Randall,
  ``The R axion from dynamical supersymmetry breaking,''
  Nucl.\ Phys.\  B {\bf 426}, 3 (1994)
  [arXiv:hep-ph/9405345].
}
\lref\DimopoulosWW{
  S.~Dimopoulos, G.~R.~Dvali, R.~Rattazzi and G.~F.~Giudice,
  ``Dynamical soft terms with unbroken supersymmetry,''
  Nucl.\ Phys.\ B {\bf 510}, 12 (1998)
  [arXiv:hep-ph/9705307].
}
\lref\GiudiceNI{
  G.~F.~Giudice and R.~Rattazzi,
  ``Extracting supersymmetry-breaking effects from wave-function
  renormalization,''
  Nucl.\ Phys.\ B {\bf 511}, 25 (1998)
  [arXiv:hep-ph/9706540].
}
\lref\SWii{
  N.~Seiberg and E.~Witten,
  ``Monopoles, duality and chiral symmetry breaking in N=2 supersymmetric
  QCD,''
  Nucl.\ Phys.\ B {\bf 431}, 484 (1994)
  [arXiv:hep-th/9408099].
}
\lref\GGRR{
  G.~F.~Giudice and R.~Rattazzi,
  ``Extracting supersymmetry-breaking effects from wave-function
  renormalization,''
  Nucl.\ Phys.\ B {\bf 511}, 25 (1998)
  [arXiv:hep-ph/9706540].
}
\lref\AFGJ{D.~Anselmi, D.~Z.~Freedman, M.~T.~Grisaru and A.~A.~Johansen,
``Nonperturbative formulas for central functions of supersymmetric gauge
theories,''
Nucl.\ Phys.\ B {\bf 526}, 543 (1998)
[arXiv:hep-th/9708042].
}
\lref\BaggerHH{
  J.~Bagger, E.~Poppitz and L.~Randall,
  ``The R axion from dynamical supersymmetry breaking,''
  Nucl.\ Phys.\ B {\bf 426}, 3 (1994)
  [arXiv:hep-ph/9405345].
}
\lref\ColemanPY{
  S.~R.~Coleman,
  ``The Fate Of The False Vacuum. 1. Semiclassical Theory,''
  Phys.\ Rev.\ D {\bf 15}, 2929 (1977)
  [Erratum-ibid.\ D {\bf 16}, 1248 (1977)].
}
\lref\LutyVR{
  M.~A.~Luty and J.~Terning,
  ``Improved single sector supersymmetry breaking,''
  Phys.\ Rev.\ D {\bf 62}, 075006 (2000)
  [arXiv:hep-ph/9812290].
}
\lref\AmaritiVK{
  A.~Amariti, L.~Girardello and A.~Mariotti,
  ``Non-supersymmetric Metastable vacua in SU(N) SQCD with adjoint matter,''
  arXiv:hep-th/0608063.
}
\lref\BanksDF{
  T.~Banks,
  ``Cosmological supersymmetry breaking and the power of the pentagon: A model
  of low energy particle physics,''
  arXiv:hep-ph/0510159.
}

\lref\DineAG{
  M.~Dine, A.~E.~Nelson, Y.~Nir and Y.~Shirman,
  ``New tools for low-energy dynamical supersymmetry breaking,''
  Phys.\ Rev.\ D {\bf 53}, 2658 (1996)
  [arXiv:hep-ph/9507378].
}

\lref\CheungIJ{
  C.~Cheung and J.~Thaler,
  JHEP {\bf 0608}, 016 (2006)
  [arXiv:hep-ph/0604259].
}
\lref\IT{
  K.~A.~Intriligator and S.~D.~Thomas,
  ``Dynamical Supersymmetry Breaking on Quantum Moduli Spaces,''
  Nucl.\ Phys.\  B {\bf 473}, 121 (1996)
  [arXiv:hep-th/9603158].
}
\lref\AffleckUZ{
  I.~Affleck, M.~Dine and N.~Seiberg,
  ``Calculable Nonperturbative Supersymmetry Breaking,''
  Phys.\ Rev.\ Lett.\  {\bf 52}, 1677 (1984).
}

\lref\IY{
  K.~I.~Izawa and T.~Yanagida,
  ``Dynamical Supersymmetry Breaking in Vector-like Gauge Theories,''
  Prog.\ Theor.\ Phys.\  {\bf 95}, 829 (1996)
  [arXiv:hep-th/9602180].
}
\lref\Chacko{
  Z.~Chacko, M.~A.~Luty and E.~Ponton,
  ``Calculable dynamical supersymmetry breaking on deformed moduli spaces,''
  JHEP {\bf 9812}, 016 (1998)
  [arXiv:hep-th/9810253].
}
\lref\IT{
  K.~Intriligator and S.~D.~Thomas,
  ``Dynamical Supersymmetry Breaking on Quantum Moduli Spaces,''
  Nucl.\ Phys.\ B {\bf 473}, 121 (1996)
  [arXiv:hep-th/9603158].
}
\lref\BarnesZN{
  E.~Barnes, K.~Intriligator, B.~Wecht and J.~Wright,
  ``N = 1 RG flows, product groups, and a-maximization,''
  Nucl.\ Phys.\ B {\bf 716}, 33 (2005)
  [arXiv:hep-th/0502049].
}
\lref\FayetJB{
  P.~Fayet and J.~Iliopoulos,
 ``Spontaneously broken supergauge symmetries and Goldstone
  spinors,''
  Phys.\ Lett.\ B {\bf 51}, 461 (1974).
}
\lref\MurayamaYF{
  H.~Murayama and Y.~Nomura,
  ``Gauge mediation simplified,''
  arXiv:hep-ph/0612186.
}
\lref\KitanoXG{
  R.~Kitano, H.~Ooguri and Y.~Ookouchi,
  ``Direct mediation of Metastable supersymmetry breaking,''
  Phys.\ Rev.\  D {\bf 75}, 045022 (2007)
  [arXiv:hep-ph/0612139].
}
\lref\CsakiWI{
  C.~Csaki, Y.~Shirman and J.~Terning,
  ``A simple model of low-scale direct gauge mediation,''
  arXiv:hep-ph/0612241.
}
\lref\AmaritiVK{
  A.~Amariti, L.~Girardello and A.~Mariotti,
  ``Non-supersymmetric Metastable vacua in SU(N) SQCD with adjoint matter,''
  JHEP {\bf 0612}, 058 (2006)
  [arXiv:hep-th/0608063].
}
\lref\LeighSJ{
  R.~G.~Leigh, L.~Randall and R.~Rattazzi,
  ``Unity of supersymmetry breaking models,''
  Nucl.\ Phys.\ B {\bf 501}, 375 (1997)
  [arXiv:hep-ph/9704246].
}
\lref\MurayamaPB{
  H.~Murayama,
  ``A model of direct gauge mediation,''
  Phys.\ Rev.\ Lett.\  {\bf 79}, 18 (1997)
  [arXiv:hep-ph/9705271].
}
\lref\CachazoJY{
  F.~Cachazo, K.~A.~Intriligator and C.~Vafa,
  ``A large N duality via a geometric transition,''
  Nucl.\ Phys.\ B {\bf 603}, 3 (2001)
  [arXiv:hep-th/0103067].
}
\lref\primer{S.~P.~Martin,
  ``A supersymmetry primer,''
  arXiv:hep-ph/9709356.
}
\lref\DuncanAI{
  M.~J.~Duncan and L.~G.~Jensen,
  ``Exact tunneling solutions in scalar field theory,''
  Phys.\ Lett.\ B {\bf 291}, 109 (1992).
}

\lref\IntriligatorFK{
  K.~A.~Intriligator and S.~Thomas,
  ``Dual descriptions of supersymmetry breaking,''
  arXiv:hep-th/9608046.
}

\lref\SWi{
  N.~Seiberg and E.~Witten,
  ``Electric - magnetic duality, monopole condensation, and confinement in N=2
  supersymmetric Yang-Mills theory,''
  Nucl.\ Phys.\ B {\bf 426}, 19 (1994)
  [Erratum-ibid.\ B {\bf 430}, 485 (1994)]
  [arXiv:hep-th/9407087].
}

\lref\DKi{
D.~Kutasov,
``A Comment on duality in N=1 supersymmetric nonAbelian gauge
theories,''
Phys.\ Lett.\ B {\bf 351}, 230 (1995)
[arXiv:hep-th/9503086].
}

\lref\ShihAV{
  D.~Shih,
  ``Spontaneous R-Symmetry Breaking in O'Raifeartaigh Models,''
  arXiv:hep-th/0703196.
}

\lref\DKAS{
D.~Kutasov and A.~Schwimmer,
``On duality in supersymmetric Yang-Mills theory,''
Phys.\ Lett.\ B {\bf 354}, 315 (1995)
[arXiv:hep-th/9505004].
}

\lref\DKNSAS{D.~Kutasov, A.~Schwimmer and N.~Seiberg,
``Chiral Rings, Singularity Theory and Electric-Magnetic Duality,''
Nucl.\ Phys.\ B {\bf 459}, 455 (1996)
[arXiv:hep-th/9510222].
}

\lref\BrodieVX{
  J.~H.~Brodie,
  ``Duality in supersymmetric SU(N/c) gauge theory with two adjoint chiral
  superfields,''
  Nucl.\ Phys.\ B {\bf 478}, 123 (1996)
  [arXiv:hep-th/9605232].
}

\lref\PeskinGC{
  M.~E.~Peskin,
  ``The Alignment Of The Vacuum In Theories Of Technicolor,''
  Nucl.\ Phys.\ B {\bf 175}, 197 (1980).
}
\lref\DineXT{
  M.~Dine and J.~Mason,
  ``Gauge Mediation in Metastable Vacua,''
  arXiv:hep-ph/0611312.
}
\lref\PouliotSK{
  P.~Pouliot and M.~J.~Strassler,
  ``A Chiral $SU(N)$ Gauge Theory and its Non-Chiral $Spin(8)$ Dual,''
  Phys.\ Lett.\ B {\bf 370}, 76 (1996)
  [arXiv:hep-th/9510228].
}
\lref\IntriligatorNE{
  K.~A.~Intriligator and P.~Pouliot,
  ``Exact superpotentials, quantum vacua and duality in supersymmetric SP(N(c))
  gauge theories,''
  Phys.\ Lett.\ B {\bf 353}, 471 (1995)
  [arXiv:hep-th/9505006].
 }

\lref\DineVC{
  M.~Dine, A.~E.~Nelson and Y.~Shirman,
  ``Low-Energy Dynamical Supersymmetry Breaking Simplified,''
  Phys.\ Rev.\  D {\bf 51}, 1362 (1995)
  [arXiv:hep-ph/9408384].
}

\lref\IntriligatorID{
  K.~A.~Intriligator and N.~Seiberg,
  ``Duality, monopoles, dyons, confinement and oblique confinement in
  supersymmetric SO(N(c)) gauge theories,''
  Nucl.\ Phys.\ B {\bf 444}, 125 (1995)
  [arXiv:hep-th/9503179].
}
\lref\ISrev{
  K.~A.~Intriligator and N.~Seiberg,
  ``Lectures on supersymmetric gauge theories and electric-magnetic  duality,''
  Nucl.\ Phys.\ Proc.\ Suppl.\  {\bf 45BC}, 1 (1996)
  [arXiv:hep-th/9509066].
}
\lref\NelsonNF{
  A.~E.~Nelson and N.~Seiberg,
  ``R symmetry breaking versus supersymmetry breaking,''
  Nucl.\ Phys.\ B {\bf 416}, 46 (1994)
  [arXiv:hep-ph/9309299].
}
\lref\DineAG{
  M.~Dine, A.~E.~Nelson, Y.~Nir and Y.~Shirman,
  ``New tools for low-energy dynamical supersymmetry breaking,''
  Phys.\ Rev.\ D {\bf 53}, 2658 (1996)
  [arXiv:hep-ph/9507378].
}
\lref\SeibergBZ{
  N.~Seiberg,
  ``Exact results on the space of vacua of four-dimensional SUSY gauge
  theories,''
  Phys.\ Rev.\ D {\bf 49}, 6857 (1994)
  [arXiv:hep-th/9402044].
}

\lref\ISS{
  K.~Intriligator, N.~Seiberg and D.~Shih,
  ``Dynamical SUSY breaking in Metastable vacua,''
  JHEP {\bf 0604}, 021 (2006)
  [arXiv:hep-th/0602239].
}

\lref\SeibergRS{
  N.~Seiberg and E.~Witten,
  ``Electric - magnetic duality, monopole condensation, and confinement in N=2
  supersymmetric Yang-Mills theory,''
  Nucl.\ Phys.\ B {\bf 426}, 19 (1994)
  [Erratum-ibid.\ B {\bf 430}, 485 (1994)]
  [arXiv:hep-th/9407087].

ION = HEP-TH 9407087;
}

\lref\IntriligatorFK{
  K.~A.~Intriligator and S.~D.~Thomas,
  ``Dual descriptions of supersymmetry breaking,''
  arXiv:hep-th/9608046.
}
\lref\LeighSJ{
  R.~G.~Leigh, L.~Randall and R.~Rattazzi,
  ``Unity of supersymmetry breaking models,''
  Nucl.\ Phys.\ B {\bf 501}, 375 (1997)
  [arXiv:hep-ph/9704246].
}
\lref\ORaifeartaighPR{
  L.~O'Raifeartaigh,
  ``Spontaneous Symmetry Breaking For Chiral Scalar Superfields,''
  Nucl.\ Phys.\ B {\bf 96}, 331 (1975).
}
\lref\HuqUE{
  M.~Huq,
  ``On Spontaneous Breakdown Of Fermion Number Conservation And
  Supersymmetry,''
  Phys.\ Rev.\ D {\bf 14}, 3548 (1976).
}

\newbox\tmpbox\setbox\tmpbox\hbox{\abstractfont }
\Title{\vbox{\baselineskip12pt \hbox{UCSD-PTH-06-12}}}
{\vbox{\centerline{Supersymmetry Breaking, R-Symmetry
Breaking}\vskip5pt\centerline{ and Metastable Vacua}}}
\smallskip
\centerline{Kenneth Intriligator$^{1}$, Nathan Seiberg$^2$ and
David Shih$^3$}
\smallskip
\bigskip
\centerline{$^1${\it Department of Physics, University of
California, San Diego, La Jolla, CA 92093 USA}}
\medskip
\centerline{$^2${\it School of Natural Sciences, Institute for
Advanced Study, Princeton, NJ 08540 USA}}
\medskip
\centerline{$^3${\it Department of Physics, Harvard University,
Cambridge, MA 02138 USA}}
\bigskip
\vskip 1cm

\noindent Models of spontaneous supersymmetry breaking generically
have an R-symmetry, which is problematic for obtaining gaugino
masses and avoiding light R-axions. The situation is improved in
models of metastable supersymmetry breaking, which generically
have only an approximate R-symmetry.  Based on this we argue, with
mild assumptions, that metastable supersymmetry breaking is
inevitable. We also illustrate various general issues regarding
spontaneous and explicit R-symmetry breaking, using simple toy
models of supersymmetry breaking.

\bigskip

\Date{March 2007}

\newsec{Introduction}

Theories of spontaneous F-term supersymmetry breaking generically
have a global $U(1)_R$ symmetry \refs{\AffleckXZ,\NelsonNF}.  The
argument of \NelsonNF\ ties spontaneous supersymmetry breaking with
the existence of an exact R-symmetry.  This argument can easily
be extended to tie the existence of an approximate R-symmetry with
metastable supersymmetry breaking.

More explicitly, a theory with an approximate R-symmetry has a
small parameter $\epsilon$, such that for $\epsilon=0$ the theory
has an R-symmetry, but for nonzero $\epsilon$ this symmetry is
broken.  Following \NelsonNF, for $\epsilon=0$ the theory breaks
supersymmetry.   We assume that, as in all known examples,  this
happens in a compact space of vacua. Now we turn on a small but
nonzero $\epsilon$.  Clearly, the small effects of nonzero
$\epsilon$ cannot ruin the supersymmetry breaking ground states.
All they can do is slightly deform the expectation values in these
states.  However, because the theory with nonzero $\epsilon$ does not
have an R-symmetry, it follows from \NelsonNF\ that it must have supersymmetric
ground states.  For small nonzero $\epsilon$, these supersymmetric ground states are at field expectation values of order an inverse power of $\epsilon$.  Therefore, we
conclude that for nonzero but small $\epsilon$ supersymmetry is
broken in a metastable state.  The longevity of this state is
guaranteed for small $\epsilon$ because then the supersymmetric
vacua are very far in field space and the tunneling to them is
suppressed.  For a recent review of such models and for an
extensive list of references, see \IntriligatorCP.

Now, let us turn to very basic phenomenological constraints. In
order to have nonzero Majorana gaugino masses, the R-symmetry
should be broken.  This breaking can be either explicit, or
spontaneous, or both explicit and spontaneous.

One possibility is that the theory has an exact $U(1)_R$ symmetry
which is spontaneously broken at the scale of supersymmetry
breaking; this occurs, for example, in the $(3,2)$ model \AffleckXZ.  This option leads to a massless Goldstone boson -- an
R-axion -- in the spectrum, which is experimentally ruled out.
This R-axion can acquire a mass if the R-symmetry is explicitly
broken.  Indeed, in any theory of gravity we expect that there are
no global continuous symmetries; therefore high dimension
operators, whose coefficients are $\CO(1/M_{p})$, will explicitly break the
symmetry.  A specific example of such an operator arises from the
constant term in the superpotential which is needed in order to
set the cosmological constant to zero (or a very small value)
\BaggerHH.  Depending on the details of the scales of
supersymmetry and R-symmetry breaking, this contribution to the
R-axion mass might or might not be sufficient to be compatible
with the various experimental constraints.

Here, we will not consider scenarios where the $U(1)_R$ symmetry
breaking is entirely due to gravity. If we decouple gravity, then
we must include explicit R-breaking terms in the field theory.
Then, the argument above shows that the supersymmetry breaking
ground state must be metastable.  We conclude that, with some mild
assumptions, low energy supersymmetry breaking requires that we
live in a metastable state! This general observation is consistent
with the fact that all known realistic models of supersymmetry
breaking which do not involve gravity (starting with the seminal
work of \DineYW) lead to a metastable state.  We see that this
fact is not just an embarrassing nuisance, which perhaps can be
avoided with more ingenious model building -- instead,
metastability is inevitable.

Accepting metastability, even in the supersymmetry breaking sector
of the theory, is helpful because it makes it much easier to
construct models, as seen e.g.\ in \refs{\DimopoulosWW}.  In fact,
the recent work of \ISS\ suggests that metastable supersymmetry
breaking is generic in field theory and string theory.  Variants,
and other models with metastable vacua have been presented, e.g.\
in
\refs{\DineGM\DineXT\AharonyMY\AmaritiVK\KitanoXG\MurayamaYF-\CsakiWI}.

In this paper, we illustrate these and other issues using toy
models of both explicit and spontaneous R-symmetry breaking. Our
examples are all based on variants of the O'Raifeartaigh model,
which is the simplest example of renormalizable spontaneous F-term
supersymmetry breaking. Recall that the original model
\ORaifeartaighPR\ has three chiral superfields, $X$, $\phi _1$,
and $\phi _2$, with canonical K\"ahler potential $K=K_{can}=X\bar
X+\phi _1\bar \phi _1+\phi _2\bar \phi _2$, and superpotential
\eqn\oraif{W_{O'R}=\half h X \phi _1^2+ m\phi _1\phi _2 +f X.}
This theory has a $U(1)_R$ symmetry, with  $R(X)=R(\phi _2)=2$ and
$R(\phi _1)=0$. It has a classical pseudo-moduli space of
supersymmetry breaking vacua, with arbitrary $\ev{X}$. At
one-loop, this degeneracy is lifted \HuqUE, and the vacuum is at
the origin of the pseudo-moduli space,  with supersymmetry broken,
and the $U(1)_R$ symmetry unbroken. On general grounds \NelsonNF,
adding a {\it generic}, R-symmetry breaking operator
\eqn\genericop{
\delta W = \epsilon f(\phi_1,\phi_2,X)
}
to the superpotential \oraif\ will restore supersymmetry. However,
if the coefficient $\epsilon$ of that operator is small, we expect
the local analysis of the supersymmetry-breaking vacuum to be
unaffected. Moreover, we expect the lifetime of the now
metastable vacuum to be parametrically large in the small
$\epsilon$ limit. In our first example, we illustrate this general
connection between approximate R-symmetry and metastable
supersymmetry breaking by taking $\delta W =\half\epsilon
\phi_2^2$ and analyzing the resulting vacuum structure in
detail.\foot{A similar toy model was considered in
\DineGM.}

For our second example, we will analyze a class of toy models for
spontaneous R-symmetry breaking, also based on O'Raifeartaigh
models. Because the pseudo-moduli $X$ of the tree-level theory
have $R(X)=2$, the R-symmetry will be spontaneously broken if the
quantum effective potential for the pseudo-moduli has a minimum
away from the origin, $\ev{X}\neq 0$. To achieve this, we
generalize \oraif\ to include a symmetry $G$. For simplicity, we
will focus on $G=SO(N)$, with the fields $\phi_i$ in the
fundamental representation. (The qualitative features of our
analysis should carry over easily to other groups and
representations.) Weakly gauging $G$ can lead to a vacuum with
$\ev{X}\ne0$, provided that $X$ couples to the gauge group, and
the gauge group is spontaneously broken.\foot{It is also possible
to spontaneously break R-symmetry in O'Raifeartaigh models without
gauge interactions \ShihAV. This interesting scenario will not be
considered here.}

The fact that a generalized O'Rafeartaigh model, augmented with a
gauge symmetry $G$, can lead to $\ev{X}\neq 0$ has been known for
decades \WittenIH. In the original example of \WittenIH, the idea
was to use gauge interactions to generate an ``inverted hierarchy"
with $\ev{X}$ exponentially larger than the scales in the
tree-level potential. In this limit, the leading contribution to
the effective potential for the pseudo-modulus $X$ is
related to its anomalous dimension:
\eqn\vonefar{V_{eff}(X) \approx
\gamma \log\left({|X|^2\over M_{cutoff}^2}\right) V_{0}\qquad
\hbox{for $X$ large},
 }
where $V_0$ is the tree-level vacuum energy along the
pseudo-moduli space. When several fields have non-zero F-terms on
the pseudo-moduli space, $\gamma$ is the anomalous dimension of a
particular linear combination of them, as we will illustrate later
in examples.  In general, $\gamma$ is a combination of the gauge
and Yukawa couplings, given at one loop by
\eqn\gammaXgen{
\gamma ^{(1)}=c_hh^2-c_gg^2
}
with $c_h$ and $c_g$ positive numbers.  If the pseudo-modulus is
charged under the gauge group, then $c_g\ne 0$ and there is a
negative contribution to the one-loop anomalous dimension coming
from the gauge interactions. If the ratio $g/h$ is sufficiently
large, $\gamma $ is negative, and the potential slopes down at
large $X$. This is the situation considered in \WittenIH. Whether
$X$ is stabilized at a finite value, as opposed to having a
runaway $X\to \infty$, hinges on the higher loop contributions to
\vonefar. This was analyzed e.g.\ in
\refs{\DimopoulosGM\BanksMG-\KaplunovskyYX}, where it was found
that $g/h$ needs to be sufficiently large to have $\gamma<0$, but
not too large in order to avoid runaway. In this phase, the
spontaneous R-symmetry breaking occurs at a scale $\ev{X}$ which
is exponentially larger than the supersymmetry breaking scale.

Motivated by an interest in low-scale gauge mediation, we will
focus the analysis of our models on a different phase of the
theory, where $\gamma$ is positive, yet $\ev{X}$ is stabilized
at small, but nonzero values. The existence of such a
``non-hierarchical" phase depends on the details of the full
Coleman-Weinberg effective potential,
\eqn\CWgen{
 V^{(1)}_{eff} = {1\over
 64\pi^2}{\rm STr}\,\CM^4\log{\CM^2\over M_{cutoff} ^2}.}
(where ${\cal M}$ are the classical, pseudo-moduli-dependent
masses), and thus it is not at all guaranteed. As a result, this
non-hierarchical phase has not been much discussed in the
literature. (It was considered long ago in an early model building
attempt \NappiHM, and it was also discussed recently in
\refs{\DineXT,\CsakiWI}.) Here we will present a broad class of
examples of this phenomenon, and attempt the first systematic
analysis of it.

More specifically, we will show that in our gauged $SO(N)$ models,
the potential can have a minimum at small or intermediate
$\ev{X}\ne 0$, but only if the ratio $g/h$ takes values in a {\it
small window} which is typically of size $\lesssim 0.1$. On
general grounds, such a window, if it exists at all, cannot be too
large, since for sufficiently large $g/h$ the theory is in the inverted
hierarchy phase (or has runaway), while for small $g/h$ the Yukawa
interactions dominate and the $U(1)_R$ remains unbroken.

In addition, we will generalize our models to include the case
where only an $SO(n)$ subgroup of $SO(N)$ is gauged. Now the phase
structure is more intricate, and depending on the couplings the
symmetry group $SO(n)\times SO(N-n)$ can either be broken to
$SO(n-1)\times SO(N-n)$ or $SO(n)\times SO(N-n-1)$ along the
pseudo-moduli space.\foot{The question of dynamical
vacuum alignment \refs{\PreskillMZ, \PeskinGC}\ does not arise in
this setup, because there are relevant interactions, which have no
reason to respect the $SO(N)$ symmetry.  In particular, the
couplings in the superpotential
will only respect the $SO(n)\times SO(N-n)$ symmetry.
The symmetry breaking pattern in the vacuum is then determined
by the tree-level values of the superpotential couplings, rather than
dynamically.}  In the
former case, the potential depends on $g$ and there can be a
minimum with $\ev{X}\ne 0$, while in the latter case the potential
is independent of $g$ and the minimum is at $\ev{X}=0$. The reason
is because, as mentioned above, the gauge group must be
spontaneously broken along the pseudo-moduli space in order for
the potential \CWgen\ to lead to $\ev{X}\neq 0$. This can be seen
from the standard expressions for the classical mass matrices
${\cal M}$, which enter in \CWgen. Because we consider situations
without $D$-term supersymmetry breaking, the masses ${\cal M}$
only depend on the gauge coupling $g$ if the gauge group is
Higgsed. If the gauge group is not Higgsed, then the potential
\CWgen\ coincides with that for $g=0$, and then the minimum of the
potential is at $\ev{X}=0$.

The outline of the paper is as follows. In section 2, we collect
some general facts and formulas about O'Raifeartaigh-type models.
We focus on two different sub-classes of
O'Raifeartaigh-type models.  The first consists of straightforward
generalizations of \oraif, while the second class consists of
models with only cubic and linear superpotential interactions. The models of
\WittenIH\ and \ISS\ belong in this second class. Section 2 will also
set up the formalism that we will need for
the examples studied in later sections. In section 3, we analyze
the effects of adding a small R-symmetry breaking operator to the original
O'Raifeartaigh model \oraif. Finally, section 4 contains our
analysis of spontaneous R-symmetry breaking in the gauged $SO(N)$
and $SO(n)\subset SO(N)$ O'Raifeartaigh models.

\newsec{General Remarks on O'Raifeartaigh models}

In this section, we review some aspects of O'Raifeartaigh-type models.  A large class of such models
has $r$ fields $X_i$,  and $s$ fields $\phi _j$, with $r>s$, with a $U(1)_R$ symmetry under
which $R(X_i)=2$ and $R(\phi _j)=0$.   We take these fields to have canonical K\"ahler potential, and superpotential
\eqn\wgenora{W=\sum _{i=1}^r X_i g_i(\phi _j).}
 For generic functions $g_i$, it is impossible for all $-F_{X_i}^\dagger =g_i(\phi _j)$ to vanish, so supersymmetry is generically broken.

The tree-level potential for the scalars is
\eqn\vtreeorgenx{V_{tree}=\sum _{i=1}^r
\Big| g_i(\phi _j)\Big|^2 +\sum _{j=1}^s \Big| \sum _{i=1}^r X_i
{\partial \over \partial \phi _j}g_i(\phi _j)\Big|^2.
 }
There is a pseudo-moduli space of vacua,  given by the $r$ fields
$X_i$, subject to the $s$ conditions
\eqn\vtreepseudomod{0=\sum _{i=1}^r  X_i {\partial \over \partial \phi _j}g_i(\phi _j )\qquad\hbox{for all}\qquad j=1\dots s.}
For generic functions $g_i(\phi _j)$, this pseudo-moduli space has
complex dimension equal to $r-s$. The $X_i$ equations of motion
are automatically satisfied on this pseudo-moduli space.  The
$\phi _j$ are determined by their equations of motion,
\eqn\vtreejeom{0=\sum _{i=1}^r \overline{g_i(\phi _j)}{\partial \over \partial \phi _j}g_i(\phi _j)\qquad\hbox{for all}\qquad j=1\dots s,}
which is generally satisfied for a discrete set of values $\phi _j=\phi _j^{(n)}$, some of which are local minima of the potential (others are saddle points).  Whether or not a given solution $\phi _j=\phi _j^{(n)}$ is a local minimum can vary with the parameters in the $g_i(\phi _j)$.

To summarize, the tree-level potential generally has a
pseudo-moduli space of supersymmetry breaking vacua, labeled by
the $r-s$ complex dimensional space of expectation values of the
$X_i$, subject to \vtreepseudomod, along with some discrete
choices $\phi _j^{(n)}$ of the solutions of \vtreejeom.  The
pseudo-moduli are lifted at one loop, by the potential \CWgen.
The resulting vacua, in all these examples, always have
$\ev{X_i}=0$ as the minimum of the one-loop potential.   The
$U(1)_R$ symmetry is thus not spontaneously (nor explicitly)
broken in these examples.

  In the remainder of this section, we will discuss some particular examples in detail.  Some of these examples have non-R global symmetries, which can be spontaneously broken. The results will be useful in the later sections, where we discuss modifications, with broken $U(1)_R$ symmetry.

\subsec{The basic O'Raifeartaigh model $OR_1$}

Let us consider the simplest case, $r=2$, $s=1$ in \wgenora.  Taking $g_1(\phi )=\half h \phi ^2+f$ and $g_2(\phi)=m\phi$ gives, with a change of notation $\phi _1=\phi$ and $\phi _2=X_2$,
\eqn\oraif{W_{(O'R)_1}=\half h X \phi _1^2+ m\phi _1\phi _2 +f X.}
In addition to the $U(1)_R$ symmetry, there is a ${\Bbb Z_2}$
symmetry, under which $\phi _1$ and $\phi _2$ are odd. Note that
this model \oraif\ has two dimensionful parameters, $m$ and $f$.
They can be made naturally small if they are generated by
dimensional transmutation of some added  dynamics, e.g.\ as in
\DineGM.

The tree-level scalar potential is given by
\eqn\oraipotorig{
V_{tree} =\Big|
 \half h \phi _1^2+f \Big| ^2 + \Big| hX\phi _1 +m\phi _2 \Big| ^2+ \Big| m\phi _1\Big|^2.}
There is an $r-s=1$ complex dimensional pseudo-moduli space of
vacua, with $X$ and $\phi _2$ constrained by the single condition
\vtreepseudomod, which here gives $hX\phi _1+m\phi _2=0$. The
$\phi _1$ equation of motion \vtreejeom\ leads to two phases,
depending on the value of
\eqn\ydef{y\equiv \left| {hf\over m^2}\right|}
with a second order phase transition at $y=1$. Let us now describe
these phases in turn:

\lfm{1.} In the $y<1$ phase, the potential is minimized along a
pseudo-moduli space of supersymmetry breaking vacua, given by
\eqn\pmsylt{
 \phi_1=\phi_2=0,\qquad X\,\,\,{\rm arbitrary}.
 }
The ${\Bbb Z_2}$ symmetry is unbroken in this phase. The classical
pseudo-moduli space degeneracy is lifted at one-loop by the
Coleman-Weinberg potential \CWgen. Near $X=0$, the potential is
\eqn\Vmin{V_{eff}^{(1)}(X)=V_0+m_X^2 |X|^2 + \CO(|X|^4),}
where $V_0$ is a constant and
\eqn\mxylt{\eqalign{
 & m_X^2={1\over 32\pi ^2}\left|h^2m^2\right|f_1 (y)\cr
 & f_1(y)\equiv y^{-1}\left((1+y)^2\log (1+y)-(1-y)^2\log
 (1-y)-2y\right)
 }}
Higher loop corrections are suppressed by powers of $h^2$. Since
$f_1(y)$ is positive for all $y<1$, the potential \Vmin\ has an
$U(1)_R$-preserving minimum at $X=0$.

\lfm{2.} In the $y>1$ phase, there are two disjoint pseudo-moduli
spaces given by
\eqn\pmsygt{
 \phi _2=-{hX\over m}\phi_1,\qquad \phi_1= \pm i\sqrt{{2f\over h}(1-y^{-1})},\qquad X\,\,\,{\rm arbitrary}.
 }
The ${\Bbb Z_2}$ symmetry is spontaneously broken in this phase.
The Coleman-Weinberg potential again takes the form \Vmin, now
with
\eqn\mxygt{\eqalign{ & m_X^2 = {1\over 8\pi ^2}\left|h^2m^2\right|f_2(y)\cr
            & f_2(y) \equiv y^2\log
 y-(y-1)^2\log(y-1)-\left(y-1/2\right)\left(2\log\left(y-1/2\right)+1\right)
}}
Again, since $f_2(y)>0$ for all $y>1$, the one-loop effective
potential is minimized at $X=0$, and the $U(1)_R$ remains
unbroken.

\medskip

\subsec{A ``cubic only" model, $OR_2$}

Consider again \wgenora\ for $r=2$, $s=1$, now with  $g_1(\phi )=\half h _1\phi ^2+f$ and $g_2=\half h_2 \phi ^2$.  For convenience, we will write it again as
\eqn\oraifii{W_{OR_2}=\half h_1 X\phi _1^2+\half h_2\phi _2 \phi _1^2 +fX.}
While \oraif\ had two dimensionful parameters, the model \oraifii\ only has one, $f$.
The tree-level scalar potential is
\eqn\oraifiisp{V_{tree}=\Big| \half h_1\phi _1^2+f\Big| ^2
+\Big| h_1 X\phi _1+h_2 \phi _2 \phi _1\Big| ^2+\Big| \half h_2
\phi _1^2\Big| ^2.} The potential \oraifiisp\ has a
one-dimensional pseudo-moduli space of vacua, with supersmmetry
broken by the non-zero vacuum energy $V_{min}\neq 0$.  But the
spectrum of massive fields is supersymmetric, and thus the loop
corrections to the potential, such as \CWgen, vanish.  The reason
is that there is a unitary change of variables, which preserves
the K\"ahler potential, of the form,
\eqn\cov{\pmatrix{Y_1\cr Y_2}=U \pmatrix{X \cr \phi _2},}
with $U\in U(2)$, which can be used to take \oraifii\ into
\eqn\oraifiicov{W_{OR_2}\to \half h Y_1(\phi _1^2+f_1)+f_2Y_2.}
The theory thus decouples into two independent sectors.  The first
has two supersymmetric vacua, at $Y_1=0$, $\phi _1=\pm
i\sqrt{f_1}$.  Supersymmetry is broken because of the decoupled
$Y_2$ field, but that sector is a free field theory.  So the model
$OR_2$ is rather trivial.

\subsec{Generalizations of $OR_1$, with fields $\phi _1$ and $\phi
_2$ in representations of a group}

A generalization of \oraif, which we will use later, is to replace the fields $\phi _1$ and $\phi _2$ with representations $\bf{r_1}$ and $\bf{r_2}$ of some global symmetry group $G$.  We keep $X$ as a single field, in the singlet representation of $G$.  We take the superpotential to still be given by \oraif, with the parameters $h$, $m$, and $f$ in the singlet representation $G$.
We choose the representations $\bf{ r_1}$ and $\bf{ r_2}$ such that
\eqn\rireqi{\bf{r_1\otimes r_1\supseteq 1, \qquad r_1\otimes r_2 \supseteq 1,}}
so that the superpotential \oraif\ is $G$-invariant. (Note that
$\bf{r_1}$ and $\bf{r_2}$ can in general be reducible representations.) In
the notation of \wgenora, we have $r=1+|{\bf r_1}|$ and $s=|{\bf
r_2}|$.

For simplicity, let us consider the case $\bf{r_1}=\bf{r_2}\equiv
{\bf r}$, which is taken to be an $|\bf{r}|$ dimensional (real)
representation of $G$. The analysis is nearly identical to that of
section 2.1.  There are again two phases, depending on the value
of the parameter $y$ defined in \ydef.

\lfm{1.} When $y<1$, the
the absolute minimum of the tree-level potential occurs at
\eqn\pmsspecphasei{
\phi_1=\phi_2=0,\quad X\,\,\,{\rm arbitrary}
}
with $V_0=|F_X|^2=|f|^2$. The global symmetry $G$ is thus unbroken in this
phase.  The tree-level mass matrices factorize into $|\bf{r}|$
copies of the basic O'Raifeartaigh model \oraif, so the one-loop
potential is just $|r|$ times that of the basic model \oraif,
$V_{eff}^{(1)}(X; r)=|{\bf r}|V_{eff}^{(1)}(X)$.  In particular,
near the origin, it is of the form \Vmin\ with
\eqn\CWspecphasei{\eqalign{
 & m_X^2 = {|{\bf r}|h^2m^2\over32\pi^2}f_1(y)
}}

\lfm{2.} When $y>1$ the analog of the vacua \pmsygt\ is
\eqn\pmsspecphaseii{
\phi_2 = -{hX \over m }\phi_1,\quad
 (\phi_1^2)_{\bf 1}=-{2f\over h}(1-y^{-1}),\quad \phi_1^\dagger = -\sqrt{f^* h\over f h^*}\phi_1,
 \quad X  \,\,\,{\rm
 arbitrary},
}
where $(\phi_1^2)_{\bf 1}$ means the singlet component in \rireqi.
The value of the potential is $V_0=|F_X|^2+|F_{\phi _2}|^2$, with
$|F_X|^2=y^{-2}|f|^2$ and $|F_{\phi _2}|^2=2y^{-2}(y-1)|f|^2$.
Note that unlike in the $y<1$ phase, for fixed $X$ the solution to
\pmsspecphaseii\ is not unique or even discrete -- that is to say,
there can be an additional (compact) component to the
pseudo-moduli space of vacua parameterized by $\phi_1$ satisfying
the second and third equations in \pmsspecphaseii. For simplicity,
and since this is all we will need for the rest of the paper, we
will limit our discussion to the special class of models where the
solution to the $\phi_1$ equations in \pmsspecphaseii\ is unique
up to global symmetries. In that case, the global symmetry $G$ is
spontaneously broken in this phase, by $\ev{\phi _1}\neq 0$, to a
subgroup $H\subset G$. The scale of the $G\to H$ symmetry breaking
varies along the pseudo-moduli space, increasing with $|X|$,
because of the $\phi _2$ expectation value in \pmsspecphaseii.
Note that there is no value of $X$ for which $G$ is unbroken
because $\phi_1$ never vanishes.

\lfm{} The pseudo-moduli space is lifted at one loop, but there
is, in the full quantum theory, a compact moduli space of vacua,
the Goldstone boson manifold $G/H$, of real dimension $|G/H|$.
Decomposing $\bf{r}$ into $H$ representations, it contains $|{\bf
r}|-|G/H|$ singlets. Thus, the classical mass spectrum of the
$\phi _1$ and $\phi _2$ fields coincides with that of $|{\bf
r}|-|G/H|$ decoupled copies of the basic O'Raifeartaigh model with
$y>1$, and $|G/H|$ copies of the basic O'Raifeartaigh model with
$y=1$ (these supply the needed massless Goldstone bosons).  So the
one-loop potential \CWgen\ is
\eqn\voneorgenrepyb{V_{eff}^{(1)}(X; {\bf r}, y>1)
 =(|{\bf r}|-|G/H|)V_{eff}^{(1)}(X; y>1)+|G/H|V_{eff}^{(1)}(X; y=1).}
In particular, the minimum is at $X=0$, around which the potential
takes the form \Vmin\ with
\eqn\Veffanswer{\eqalign{
 & m_X^2 = {h^2m^2\over
8\pi^2}\left( (|{\bf r}|-|G/H|)f_2(y)+\half |G/H|(\log
4-1)\right).
}}
We stress that this analysis only applies to the special class of
models where $\phi_1$ is completely specified up to global
symmetries by the equations \pmsspecphaseii.

\bigskip

In section 4, we will reconsider these models, with the symmetry
$G$ (or a subgroup) replaced with a gauge, rather than global,
symmetry. In the $y<1$ phase, $G$ remains unbroken, so the 1-loop
potential is unaffected by the gauging and the $U(1)_R$ symmetry
remains unbroken. (It is amusing to note that, even if $U(1)_R$
were to be spontaneously broken, the messenger mass matrix $\CM$
in all of these models satisfies $\det \CM=const$ \ShihAV, and so
the gauginos would remain massless to leading order in the SUSY
breaking.) On the other hand, in the $y>1$ phase $G$ is broken, so
here the 1-loop potential is affected by the gauging and can have
a minimum away from the origin.

\subsec{Generalizations of $OR_2$, with fields $\phi _1$ and $\phi
_2$ representations of a group}

We noted in section 2.2 that the model with superpotential
\oraifii\ was rather trivial, because it decoupled into a
supersymmetric sector and a free field theory.  We now consider
generalizing the model, with the same superpotential \oraifii, by
making $\phi _1$ and $\phi _2$ representations $\bf r_1$ and $\bf
r_2$ of a global symmetry group $G$. Again, we keep $X$ as a
singlet representation of $G$. The couplings $h_1$, $h_2$, and $f$
are singlets of $G$, and the superpotential \oraifii\ is $G$
invariant if the representations satisfy
\eqn\rireqii{\bf{r_1\otimes r_1\supseteq 1\oplus \overline{r_2}}.}
When ${\bf r_2}\neq {\bf 1}$,  there is no  longer a unitary
change of variables generalizing \cov, to bring the theory to a
decoupled form analogous to \oraifiicov.  So this modified theory
is nontrivial.

These models are a particular example of the general class of models \wgenora, with $r=1+|{\bf r_2}|$
and $s=|{\bf r_1}|$.  If the superpotentials were generic, and unrestricted by the  $G$ symmetry, we would then have that supersymmetry is broken if  $r>s$, i.e. if $|{\bf r_2}|\geq |{\bf r_1}|$, and
unbroken otherwise.  Let us consider this in more detail, accounting for the particular, $G$ symmetric, form of the superpotential.   The tree-level potential is
\eqn\vtreeoriig{V_{tree}=\Big| \half h_1(\phi _1^2)_{\bf 1}+f\Big| ^2
 +\Big| h_1 X \phi _1 +h_2 (\phi _2\phi _1)_{\bf r_1}\Big| ^2
 +\Big| \half h_2 (\phi _1^2) _{\bf \overline{r_2}}\Big|^2,}
where $(\phi _1^2)_{\bf 1}$ and  $(\phi _1^2)_{\bf \overline{r_2}}$ are the singlet and the ${\bf \overline{r_2}}$ representations in \rireqii, respectively.  We can always choose the $X$ and $\phi _2$ expectation values such that the middle term in \vtreeoriig\ vanishes,
\eqn\oriigpms{h_1 X\phi _1 +h_2(\phi _2\phi _1)_{\bf r_1}=0.}
This is $|{\bf r_1}|$ conditions on the $1+|{\bf r_2}|$ fields $X$ and $\phi _2$, so \oriigpms\ is satisfied
on a (pseudo)moduli space of superficial dimension $|{\bf r _2}|+1-|{\bf r_1}|$ (the actual dimension can differ from that).  These models will break supersymmetry iff there is no simultaneous solution to
\eqn\breaksusycond{(\phi _1^2)_{\bf 1}\neq 0, \qquad (\phi _1^2)_{\bf \overline{r_2}}=0.}
Whether or not there are supersymmetric vacua, where
\breaksusycond\ can be satisfied, depends on the representations
$\bf r_1$ and $\bf r_2$.

The $X$ and $\phi _2$ equations of motion are satisfied on the
space \oriigpms.  The $\phi _1$ equations of motion are cubic, and
one solution is always $\ev{\phi _1}=0$. However, this is always a
saddle point of the potential, as seen by expanding \vtreeoriig\
to quadratic order around $\phi _1=0$, and noting that there is a
tachyonic mode, where the first term in \vtreeoriig\ can be
reduced without affecting the other terms. The minima of the
potential are given by the non-zero solutions, $\phi _1=\phi
_1^0\neq 0$. Expanding around such a solution, the symmetry $G$ is
broken to some subgroup $H$. This subgroup will in general depend
on $\phi_1^0$; in particular, it will be enhanced at special
values of $\phi_1^0$. It is generally the case that the points of
enhanced unbroken symmetry are extrema of the effective potential,
and in many  (if not all) examples, the point of maximal unbroken
global symmetry is a local minimum.

Let us consider some examples.  Take $G=SU(N)$ and let ${\bf r_1}$ and $\bf r_2$ be the adjoint representation.  This is a model considered in \WittenIH\ (with $G=SU(5)$ gauged, and identified with the GUT gauge group).  In this case, there is no solution of \breaksusycond, so supersymmetry is broken.
Another example is $G=SU(N_f)\times SU(N)$, with $\bf r_1=(N_f,
N)\oplus (\bar{N_f}, \bar N)$ and $\bf r_2=(N_f^2-1, 1).$  This
example is the supersymmetry breaking model analyzed in \ISS, with
$\phi _1=\varphi\oplus\widetilde \varphi $, $\phi _2=\Phi -
{1\over N_f}\Tr ~\Phi$, and $X=\Tr ~\Phi$. This model breaks
supersymmetry if $N_f>N$, and in this model, the general condition
\breaksusycond\ is the ``rank condition" of \ISS.

To summarize, in this class of ``cubic only" models, there is a
pseudo-moduli space given by the solutions to \oriigpms\ and to
the $\phi_1$ equations of motion. Along this pseudo-moduli space,
the global symmetry $G$ is always spontaneously broken by $\phi_1
=\phi _1^0\neq 0$ to some subgroup $H$, for all values of the
parameters in the superpotential. Therefore, regardless of where
the effective potential is minimized on this pseudo-moduli space,
there is always a compact moduli space of vacua, a Goldstone boson
manifold $G/H$. This should be contrasted with the $OR_1$ class of
models discussed in the previous subsection, where the symmetry
breaking phase depended on the parameter $y=h f/m^2$.

\newsec{Metastable SUSY Breaking in a Modified O'Raifeartaigh
Model}

In section 2, we saw that for all values of the couplings, the
vacuum of the basic O'Raifeartaigh model occured at $\ev{X}=0$,
where $U(1)_R$ is unbroken. Let us now consider what happens when
a small, explicit R-symmetry breaking operator is added to the
superpotential \oraif. Generically, the presence of such an
operator introduces SUSY vacua elsewhere in field space,
rendering the SUSY-breaking vacuum metastable. However, as long
as the coupling is small, these vacua will be well separated and
the SUSY-breaking vacuum parametrically long-lived.

To be concrete, let us consider the superpotential
\eqn\worm{W=\half hX \phi _1^2+ m\phi _1\phi _2 +fX + \half
 \epsilon m\phi_2^2}
with $|\epsilon | \ll 1$.  The classical scalar potential is now
 \eqn\orpotm{V_{tree}=\Big|
 \half h \phi _1^2+f \Big| ^2 + \Big| hX\phi _1 +m\phi _2 \Big| ^2+ \Big| m\phi _1
 + \epsilon m \phi_2\Big|^2.}
There are two supersymmetric vacua, at
 \eqn\SUSYmo{\ev{\phi_1}_{susy}=\pm \sqrt{-2f/h} ,
 \qquad \ev{\phi_2}_{susy}=  \mp{1
 \over \epsilon} \sqrt{-2f/h}, \qquad \ev{X}_{susy}=
 { m \over h\epsilon}. }
 For small $\epsilon$, the supersymmetric vacua \SUSYmo\ have $X$ far from
the origin.  As $\epsilon \to 0$, these supersymmetric vacua are
pushed to infinity.

In addition to these supersymmetric vacua, the scalar potential
\orpotm\ is approximately minimized along the pseudo-moduli spaces
of the two phases. For $y<1$, the pseudo-moduli space of the
$\epsilon=0$ theory \pmsylt\ remains an extremum of the potential
of the $\epsilon\ne 0$ theory. The classical masses of the
fermionic and scalar components of $\phi _1$ and $\phi _2$ around
\pmsylt\ are given by
 \eqn\eigenii{\eqalign{ m_{0}^2 &=\half\Big(
 |hX|^2+|m|^2(2+|\epsilon|^2)+\eta|hf| \cr
 & \qquad \pm\sqrt{(|hX|^2+|m|^2(2+|\epsilon|^2)+\eta|hf|)^2 - 4m^2(|h X
\epsilon-m|^2+\eta |hf|(1+|\epsilon|^2))}\Big)\cr
 m_{1/2}^2 &= m_0^2\big|_{f=0}}}
where $\eta =\pm 1$. In order for the pseudo-moduli space of
supersymmetry breaking vacua to be locally stable, without
tachyonic modes, the eigenvalues \eigenii\ must all be positive.
There are indeed no tachyonic modes for a range of the
pseudo-modulus $X$:
\eqn\nontapm{\left|1-{\epsilon h X\over m}\right|^2  >
 \left( 1+|\epsilon|^2 \right) y}
Outside of the range \nontapm,  there is one tachyonic mode in
\eigenii, and the pseudo-moduli space is there locally unstable,
as there the fields can roll down the tachyonic direction to the
supersymmetric vacua \SUSYmo.    In the region where \nontapm\ is
satisfied, the pesudomoduli space is locally stable. Note that
this region includes a large neighborhood of the origin $X=0$ for
all
\eqn\ylonerange{y<1/(1+|\epsilon|^2).}

\
The classical pseudo-moduli space degeneracy is lifted by the
1-loop effective potential, which we compute using \CWgen, with
the masses \eigenii.  We restrict our attention to the range
\nontapm\ and \ylonerange,  where the pseudo-moduli space is
locally stable. The effective potential $V_{eff}(X)$ thus computed
is found to have a minimum near the origin.  In particular,
expanding near the origin, the result is
\eqn\CWeps{
V^{(1)}_{CW}=V_0 +  m_X^2|X-X_{min}|^2
+\CO(\epsilon^2,|X-X_{min}|^4)
}
To order $\epsilon$, the local minimum is moved from the origin to
\eqn\Xeps{X_{min}= -{\epsilon
m\over
h}{(1+y)\log(1+y)-(1-y)\log(1-y)-2y\over(1+y)^2\log(1+y)-(1-y)^2\log(1-y)-
2y}+\CO(\epsilon^3).} This is a metastable vacuum, with
supersymmetry broken.
The light spectrum in this vacuum consists of the massless
Goldstone fermion, $\psi _X$. All other modes have ${\cal
O}(\epsilon ^0)$ masses.  In particular, the pseudo-modulus $X$
has mass $m_X$ as in \mxylt, and there is  no light (pseudo)
Goldstone boson ``R-axion."

For $y>1$, the pseudo-moduli space of the $\epsilon=0$ theory
\pmsygt\ is no longer an exact extremum of the potential; instead,
there is an $\CO(\epsilon)$ tadpole for $X$. However, we expect
that for $\epsilon\ll h$, this tadpole will be stabilized by a
positive $m_X^2$ at one-loop, approximately \mxygt\ to leading
order in $\epsilon$. To make this more precise, let us integrate
out $\phi_1$ and $\phi_2$ exactly at tree-level. Their
mass-squareds are complicated functions of $h$, $X$, $m$, $f$ and
$\epsilon$, analogous to \eigenii. Curiously, they are positive as
long as the {\it reverse} of \nontapm\ is satisfied:
\eqn\nontapmrev{\left|1-{\epsilon h X\over m}\right|^2  <
 \left( 1+|\epsilon|^2 \right)y}
For $y>1$, this inequality always includes a neighborhood around
$X=0$. Integrating out $\phi_i$, we obtain an effective potential
for $X$, which is an expansion in powers of $\epsilon X$,
$\epsilon X^*$:
\eqn\VeffX{
V_{eff} = -{m^3\epsilon(y-1)\over32\pi^2 h}(X+X^*) +
\CO(\epsilon^2 X^2,\, \epsilon^2 (X^*)^2,\, \epsilon^2|X|^2)
}
Adding to this the one-loop potential \Vmin\mxygt, we see that
there is indeed a local minimum at
\eqn\Xminygt{
X_{min} ={\epsilon m(y-1)\over h^3 f_2(y) }(1+\CO(h^2))
}
Here the $\CO(h^2)$ corrections include not only the corrections
to $m_X^2$, but also the $\CO(\epsilon h)$ tadpole term that is
expected to appear in the one-loop Coleman-Weinberg potential, due
to the explicitly broken R-symmetry.

Finally, let us consider the lifetime of these metastable
SUSY-breaking vacua. For $\epsilon \ll 1$, the metastable vacuum
\Xeps\Xminygt\ is widely separated from the supersymmetric vacua
\SUSYmo\ by an $\CO(1/\epsilon)$ distance in field space. This
ensures that the supersymmetry breaking vacuum is parametrically
long lived as $\epsilon \to 0$. In particular, the bounce action
scales as $S_{bounce}\sim \epsilon ^{-\alpha}$ for some
$\alpha>0$, and can be made arbitrary large for $\epsilon $
sufficiently small.

\newsec{Spontaneous $U(1)_R$ Breaking in a gauged $SO(N)$ Model}

Having analyzed a toy model with explicit $U(1)_R$ breaking, we
now turn to a model with spontaneous $U(1)_R$ breaking. As
described in the introduction, this can be achieved by gauging a
global symmetry $G$ in an O'Raifeartaigh type model. If $G$ is
spontaneously broken along the pseudo-moduli space, then for
appropriate values of the parameters, the pseudo-modulus can get a
negative mass-squared around the origin.

Specifically, the models we will consider in this section are
those of section 2.3, with $G=SO(N)$ and the fields $\phi_{1,2}$
transforming in the fundamental representation.\foot{For
$N=6$, in the $y\to \infty$ limit, this model reduces to the IR
description of the ITTY model \refs{\IT,\IY}.} Thus, our superpotential is given by
\eqn\WSON{
W = {1\over2}hX\vec\phi_1^2+m\vec\phi_1\cdot\vec\phi_2 + f X
}
For $N=2$ this model was studied recently in \DineXT. For
simplicity, we will take all the couplings to be real and positive
throughout this section, which can be always be done via field
phase rotations. In section 2.3, we saw that without gauging the
global symmetry, the model always has a $U(1)_R$ preserving vacuum
at $X=0$. Now let us analyze what happens with we gauge the full
$SO(N)$ global symmetry. This introduces D-terms into the scalar
potential, $V_D={1\over2}g^2(D^{(ab)})^2$ where
\eqn\Dtermspec{ D^{(ab)}=\phi_1^\dagger
T^{(ab)}\phi_1+\phi_2^\dagger T^{(ab)}\phi_2
 }
and $T^{(ab)}$ are the generators of $SO(N)$,
\eqn\SONgenerators{
T^{(ab)}\,_{cd} = \delta^a_c\delta^b_d-\delta^a_d\delta^b_c
}
with $1\le a<b\le N$. Substituting \pmsspecphasei\ and
\pmsspecphaseii\ into \Dtermspec, we see that the D-terms vanish
identically along the pseudo-moduli spaces of both the $y<1$ and
the $y>1$ phases. Therefore, \pmsspecphasei, \pmsspecphaseii\
remain the absolute minima of the tree-level scalar potential in
their respective phases, even after gauging the $SO(N)$.

In the $y<1$ phase, the $SO(N)$ symmetry is unbroken, so the
one-loop effective potential is independent of $g$ as discussed in
the introduction. Thus the one-loop effective potential is still
given by \CWspecphasei, with an R-preserving minimum at $X=0$.

In the $y>1$ phase, however, the $SO(N)$ symmetry is spontaneously
broken, so now the potential depends on $g$. At small $|X|$, we
have instead of \Veffanswer
\eqn\CWphaseiig{
V^{(1)}_{CW} = const. + {h^2m^2\over
16\pi^2}\Big((N-1)f_3(\eta)+2f_2(y)\Big)|X|^2 + \CO(|X|^4)
}
where
\eqn\ftwodef{\eqalign{ f_3(\eta) & ={1\over  1 + 2\eta^2 }
        \Bigg(
          6\eta^4( 1 + 2\,{\eta }^2) \log 2\eta^2 + 2( 1 + \eta^4 + 2\eta^6 ) \log (2 +2\eta^2) \cr
        &\qquad\qquad\qquad -
          2( 1 + 2\eta^2 )( 1 + 4\eta^4)\log (1 + 2\eta^2) -{( 1 - 2\eta^2) }^2
          \Bigg)
          }}
and we have defined
\eqn\etadef{
\eta = {g\over h}\sqrt{2(y-1)}
}
Now since $f_2(y)$ is bounded and positive for $y>1$, while
$f_3(\eta)$ is positive at $\eta=0$ but is unbounded from below as
$\eta\to \infty$, we see that there must exist some
$\eta_{min}(y)$ such that when
\eqn\etamin{
\eta > \eta_{min}(y)
}
the pseudo-modulus is tachyonic at $|X|=0$.

Although we have shown that for $\eta>\eta_{min}$, the
R-preserving vacuum at the origin is destabilized, it remains to
be seen whether there is a vacuum for any $|X|\ne 0$, or whether
the potential simply runs away to infinity. At large $X$, the
potential \CWgen\ has the behavior
\eqn\VCWlargeY{
V_{CW}^{(1)} \propto h^2\Big( N-4(N-1)\eta^2 \Big)\log{|X|\over
M_{cutoff}} + \CO(|X|^{-1}).
}
This also follows from \vonefar.  The tree-level vacuum energy density is
$V_0=|F_X|^2+|F_{\phi _2}|^2$, with $|F_X|^2=y^{-2}|f|^2$ and $|F_{\phi _2}|^2=2y^{-2}(y-1)|f|^2$.  The renormalization of these terms leads to \vonefar\ with the anomalous dimension of the pseudo-modulus given by
\eqn\gammaexis{\gamma={|F_X|^2\gamma _X+|F_{\phi _2}|^2\gamma _{\phi _2}\over |F_X|^2+|F_{\phi _2}|^2}={\gamma _X+2(y-1)\gamma _{\phi _2}\over 2y-1},}
Here $\gamma _X$ and $\gamma _{\phi _2}$ are the anomalous dimensions of those fields, given at one loop by
\eqn\anomdimsare{\gamma _X^{(1)}={h^2N\over 32\pi ^2}, \qquad \gamma _{\phi _2}^{(1)}=-{g^2(N-1)\over 8\pi ^2},}
where the factor of $N-1$ comes from the Casimir $C_2({\bf
r})=|G|T({\bf r})/|{\bf r}|$. Using \anomdimsare\ in \gammaexis,
we thus obtain for the anomalous dimension of the pseudo-modulus
$\gamma =h^2(N-4\eta ^2(N-1))/32\pi ^2(2y-1)$.  Using this in
\vonefar\ agrees with \VCWlargeY.

So we see from \VCWlargeY\ that as long as
\eqn\etamax{
\eta<\eta_{max} = {1\over2}\sqrt{N\over N-1}
}
the potential curves up at infinity, so $\ev{X}$ cannot be too
large.   Our interest is in the window where both \etamin\ and
\etamax\ are satisfied,
\eqn\window{
\eta_{min}(y)<\eta<\eta_{max}.
}
\bigskip
\centerline{\epsfxsize=0.60\hsize\epsfbox{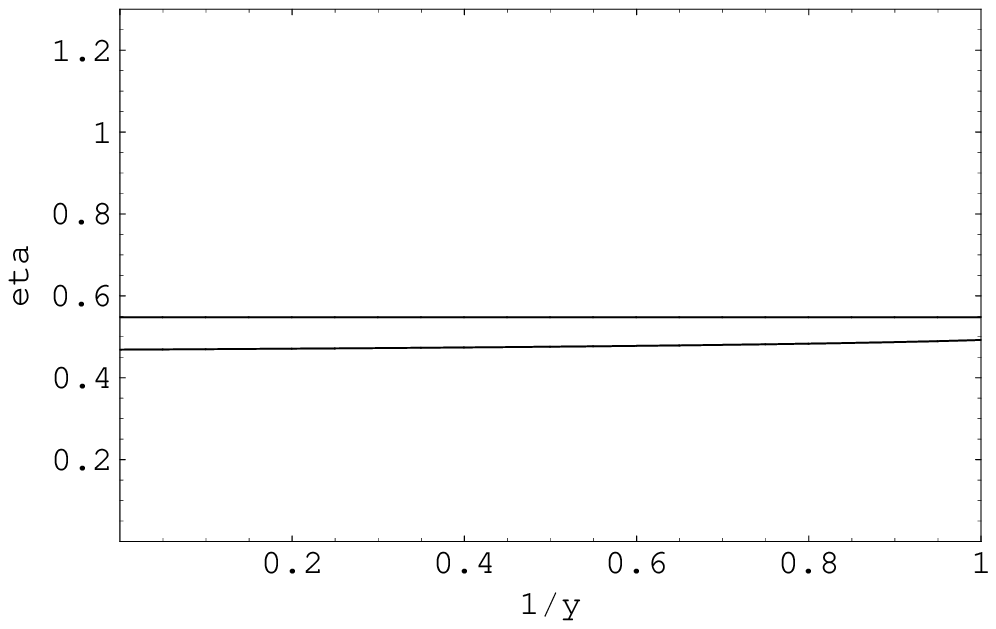}}
\noindent{\ninepoint\sl \baselineskip=8pt {\bf Figure 1}:{\sl $\;$
A plot of $\eta_{min}(y)$ (bottom curve) and $\eta_{max}$ (top
curve) vs.\ $y^{-1}$ in the gauged $SO(N=6)$ model. The window
\window\ is for $\eta$ in the region between the two solid lines.
}}
\bigskip

\noindent In this case, there is a SUSY-breaking, R-breaking
vacuum at some finite $\ev{X}$ which is not zero, and also not
hierarchically large. A plot of $\eta_{min}(y)$ and $\eta_{max}$
vs.\ $y^{-1}$ is shown in figure 1, for $N=6$. This figure also
illustrates a general feature of $\eta_{min}(y)$ -- it is a
monotonically increasing function of $y^{-1}$.

Let us make a few comments on the window \window:

\lfm{1.} The window is non-empty, for all $N$ and $y>1$. One can
verify this by, for instance, checking that $|X|$ is always
tachyonic at the origin when $\eta=\eta_{max}$.

\lfm{2.} The window is generally quite small. Figure 1 shows the
typical values of $\eta_{min}\approx 0.47$ (the dependence on $N$
is minimal), while according to \etamax, $\eta_{max}\approx
0.5-0.6$ for largish $N$. So the window in $\eta$ is typically a
size of order $\Delta \eta\approx 0.05-0.1$. In addition, for
$\eta$ close to $\eta_{max}$, the existence of the minimum of the
potential depends sensitively on the large $X$ behavior of the
potential \VCWlargeY. This is suppressed not only by the
loop-counting parameter $h^2$, but also by $\eta_{max}-\eta <
\Delta \eta \ll 1$. Since the two-loop correction to the potential
will not be suppressed by this additional factor of $\Delta\eta$,
in order for us to be able to trust the one-loop approximation, we
need $h$ (and hence $g$) to be smaller than the naive perturbation
expansion would suggest.

\lfm{3.} The existence of this window of spontaneous $U(1)_R$
breaking depends on the details of the full Coleman-Weinberg
potential, and not just its leading logarithm \VCWlargeY.
Correspondingly, the scale of $U(1)_R$ breaking $\langle X\rangle$
is not hierarchically large, but rather is $\CO(M)$ where $M$ is
some characteristic combination of the mass scales in the
superpotential.

\lfm{4.} Because the R-symmetry is spontaneously broken, the
massless spectrum includes a real scalar, the R-axion, in addition
to the Goldstino and the $SO(N-1)$ gauge fields. The $SO(N-1)$
gauginos are massless at tree-level, but they will pick up a
Majorana mass term at one-loop. (One possibility that we have not
checked in this example is that the gaugino masses vanish to
leading order in the SUSY breaking. This happens in O'Raifeartaigh
models without spontaneous gauge symmetry breaking, as we discuss
at the end of section 2.3.)

\medskip

Finally, let us discuss what happens for $\eta>\eta_{max}$. This
regime corresponds to the ``inverted hierarchy" phase first
studied by Witten in \WittenIH. Here the 1-loop potential has
runaway at infinity, and there may or may not be a vacuum at
exponentially large fields, depending on the details of the
renormalization group equations for $g$ and $h$. While the
``inverted hierarchy" and its uses for model building are
well-known (see e.g.\
\refs{\WittenIH\DimopoulosGM\BanksMG-\KaplunovskyYX}), the uses of
the non-hierarchical phase have been relatively unexplored (see
however \NappiHM, and the more recent work of
\refs{\DineXT,\CsakiWI}). However, if the non-hierarchical phase
always occurs in a small window of couplings such as \window, then
the usefulness of this phase for (natural) model building might be
limited.

\subsec{Gauged $SO(n)\subset SO(N)$ -- tree-level vacuum
alignment/mis-alignment}

In this subsection, we will analyze the $SO(N)$ model with a
subgroup $SO(n)\subset SO(N)$ gauged. As we shall see, the model
becomes much more complicated, so we will focus mainly on a few
qualitative physics points and be brief with the technical
details.

Since we have broken the $SO(N)$ global symmetry explicitly by
gauging an $SO(n)$ subgroup, with $SO(n)$ gauge coupling $g\neq 0$, we should consider the most general
$SO(n)\times SO(N-n)$ invariant superpotential of the form \oraif.
This is given by:
 \eqn\wsvyiii{W=\half h
 X \phi_1^2+\half\tilde h X \tilde\phi_1^2+m \phi_2 \cdot \phi_1 + \tilde m\tilde\phi_2\cdot
 \tilde
 \phi_1
 +f X}
where $\phi_1$, $\phi_2$ ($\tilde \phi_1$, $\tilde \phi_2$)
transform in the fundamental of $SO(n)$ ($SO(N-n)$).  Because $SO(N)$ is not a symmetry, we generally have $h\neq \tilde h$ and $m\neq \tilde m$ (taking them to be equal would not be preserved by renormalization.).   Because of the tree-level interactions, there is no limit where $SO(N)$ is restored, even as an accidental symmetry.   So there is no issue of dynamical vacuum alignment here. Whether the vacuum aligns to break, or not break, the $SO(n)$ gauge symmetry is determined entirely at tree-level, by the couplings in the superpotential  \wsvyiii. Indeed, there are now three different phases depending
on the ratios $y=|hf/m^2|$ and $\tilde y = |\tilde h f / \tilde
m^2|$ (these are summarized in figure 2):

\lfm{1.} When $y$, $\tilde y < 1$, the F and D-terms are minimized
with
\eqn\pmsphaseison{
\phi_i=\tilde\phi_i=0,\quad X\,\,\,{\rm arbitrary}
}
with $V_0=f^2$. Since the gauge symmetry is unbroken, the one-loop
Coleman-Weinberg potential is independent of the gauge coupling,
and so it reduces to an obvious generalization of \CWspecphasei:
\eqn\CWspecphaseison{
 m_X^2 = {1\over32\pi^2}\left(n
h^2m^2f_1(y)+(N-n)\tilde h^2\tilde m^2f_1(\tilde y)\right)
}
So in this phase the R-symmetry remains unbroken.

\lfm{2.} $\tilde y>1$, $y<\tilde y$. Now the scalar potential is
minimized along
\eqn\pmsphaseiison{
\phi_i=0,\qquad \tilde\phi_2= -{\tilde h X\over \tilde m
}\tilde\phi_1,\qquad \tilde\phi_1^2=-|\tilde\phi_1^2|=-{2f\over \tilde h}(1-\tilde y^{-1}),\qquad
X\,\,\,{\rm arbitrary}
}Since $\phi_i=0$, the pseudo-moduli space preserves the full
$SO(n)$ gauge symmetry. So this phase corresponds to tree-level
vacuum alignment. Since the gauge symmetry is again unbroken, the
one-loop CW potential is independent of $g$. In fact, it is a
straightforward generalization of \Veffanswer,
\eqn\Veffanswerson{\eqalign{
  m_X^2 =  {1\over
32\pi^2}\Bigg(h^2m^2 n f_1(y/\tilde y)+ \tilde h^2\tilde
m^2\Big[2(N-n-1)(\log 4-1)+4f_2(\tilde y)\Big]\Bigg)
}}
and consequently, the R-symmetry remains unbroken.

\lfm{3.} $y>1$, $\tilde y<y$:
\eqn\pmsphaseiiison{
\tilde\phi_i=0,\qquad \phi_2 = -{h X \over m} \phi_1,\qquad
\phi_1^2=-|\phi_1^2|=-{2f\over h}(1-y^{-1}),\qquad X\,\,\,{\rm
arbitrary}
}
Since $\phi_1\ne 0$, the pseudo-moduli space spontaneously breaks
the gauge symmetry from $SO(n)\to SO(n-1)$. So this phase
corresponds to tree-level vacuum misalignment. Now the one-loop
potential depends on $g$, and its form is a generalization of
\CWphaseiig:
\eqn\CWphaseiigson{
m_X^2 = {1\over
32\pi^2}\Bigg(h^2m^2\Big[2(n-1)f_3(\eta)+4f_2(y)\Big]+\tilde
h^2\tilde m^2(N-n) f_1(\tilde y/y)\Bigg),
}
The qualitative features of this potential are the same as for
\CWphaseiig. In particular, there will be a (possibly small) range
of parameters where the R-breaking vacuum exists at $X\sim
\CO(M)$, where $M$ is again some characteristic combination of the
mass scales. In addition, there will be an ``inverted hierarchy"
phase where $X$ is exponentially larger than $M$.

\bigskip

\bigskip

\centerline{\epsfxsize=0.60\hsize\epsfbox{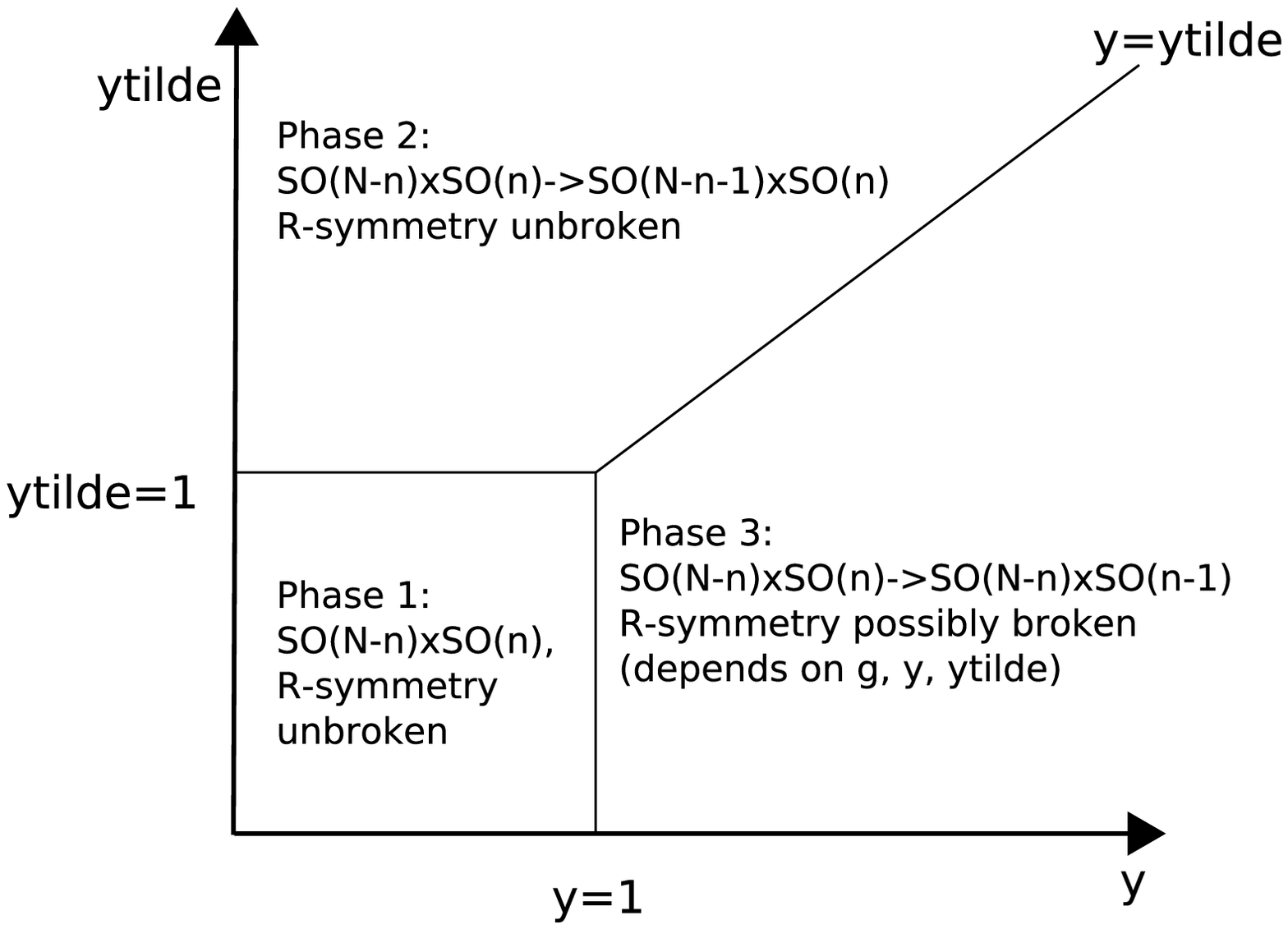}}
\noindent{\ninepoint\sl \baselineskip=8pt {\bf Figure 2}:{\sl $\;$
A phase diagram for the gauged $SO(n)\subset SO(N)$ O'Raifeartaigh
model.}}
\bigskip

\bigskip

\noindent {\bf Acknowledgments:}

We would like to thank Nima Arkani-Hamed, Michael Dine and Lisa
Randall for useful discussions. The research of NS is supported in
part by DOE grant DE-FG02-90ER40542. The research of DS is
supported in part by DOE grant DE-FG0291ER40654. The research of
KI is supported in part by UCSD grant DOE-FG03-97ER40546.  This
work was initiated when KI was at the IAS, and also supported by
IAS Einstein Fund; KI would like to thank the IAS for their
hospitality and support on his sabbatical visit.  He would also
like to thank the IHES, for hospitality and support in August
2006.

\listrefs

\end